\documentclass
[superscriptaddress,secnumarabic,amssymb,amsmath,nobibnotes,aps,prd,showkeys,showpacs,nofootinbib,onecolumn,notitlepage,12pt]{revtex4}%
\usepackage{setspace}
\usepackage{color}
\usepackage{amsmath}
\usepackage{amsfonts}
\usepackage{verbatim}
\usepackage{amssymb}
\usepackage{graphicx}
\usepackage[caption=false]{subfig}%
\providecommand{\U}[1]{\protect\rule{.1in}{.1in}}
\newcommand{\be}{\begin{equation}}
\newcommand{\ee}{\end{equation}}

\newcommand{\mincir}{\raise
-3.truept\hbox{\rlap{\hbox{$\sim$}}\raise4.truept\hbox{$<$}\ }}
\newcommand{\magcir}{\raise
-3.truept\hbox{\rlap{\hbox{$\sim$}}\raise4.truept\hbox{$>$}\ }}

\ifx\pdfoutput\relax\let\pdfoutput=\undefined\fi
\newcount\msipdfoutput
\ifx\pdfoutput\undefined\else
\ifcase\pdfoutput\else
\ifx\paperwidth\undefined\else
\ifdim\paperheight=0pt\relax\else\pdfpageheight\paperheight\fi
\ifdim\paperwidth=0pt\relax\else\pdfpagewidth\paperwidth\fi
\fi\fi\fi
\begin{document}
\title{Lie symmetries for the cosmological field equations in brane-world gravity
with bulk scalar field}
\author{Andronikos Paliathanasis}
\email{anpaliat@phys.uoa.gr}
\affiliation{Institute of Systems Science, Durban University of Technology, Durban 4000,
South Africa}
\affiliation{Departamento de Matem\'{a}ticas, Universidad Cat\'{o}lica del Norte, Avda.
Angamos 0610, Casilla 1280 Antofagasta, Chile}

\begin{abstract}
We address the group classification problem for gravitational field equations
within the context of brane-world cosmology, considering the presence of a
bulk scalar field. Our investigation revolves around a five-dimensional
spacetime, with the four-dimensional
Friedmann--Lema\^{\i}tre--Robertson--Walker geometry embedded within it.
Additionally, we assume that the scalar field exists in this five-dimensional
geometry (bulk) and possesses a nonzero mass. The resulting field equations
constitute a system of nonlinear partial differential equations. We apply the
Lie symmetry condition to identify all functional forms of the scalar field
potential, ensuring that the field equations remain invariant under
one-parameter point transformations. Consequently, we find that only the
exponential potential exhibits Lie symmetries. Finally, the Lie invariants are
used to construct similarity transformations, which enable us to derive exact
solutions for the system.

\end{abstract}
\maketitle
\date{\today}

\section{Introduction}

\label{sec1}

The theory of General Relativity is a pure geometrical theory of gravity which
states that physical space is described by a four-dimensional Riemannian
manifold \cite{ae1}. General Relativity has been extensively tested and proven
successful in explaining various gravitational phenomena \cite{grt1,grt2,grt3}%
. However, in cosmological investigations, the inclusion of the cosmological
constant $\Lambda$ becomes crucial to account for the late-time acceleration
phase of our universe \cite{rr1, Teg, Kowal}.

The cosmological constant introduces an effective fluid described by an
energy-momentum tensor featuring constant energy density $\rho_{\Lambda}$ and
pressure $p_{\Lambda}$, with a relation $p_{\Lambda}=-\rho_{\Lambda}$.
Although the cosmological constant is a very simple mechanism for the
description of the acceleration of the universe it suffers from two major
problems, the old cosmological problem and the coincidence problem
\cite{conpr1,conpr2}. The main direction followed by cosmologists to overcome
the aforementioned problems is the introduction of new degrees of freedom in
the cosmological field equations. Scalar fields \cite{q4} are particularly
relevant in this context, as they can describe the additional degrees of
freedom proposed in other models \cite{q11, q1}.

In brane-world gravity, the observable four-dimensional universe is considered
to exist within a higher-dimensional world \cite{br1, br2, br3, br4}. This
characteristic offers various mechanisms for addressing the cosmological
constant problems \cite{cc1, cc2, cc3, cc4}. Brane-world cosmological
scenarios involving bulk scalar fields have been previously studied \cite{bb0,
bb01}. In \cite{bb1}, a massless bulk scalar field was introduced, leading to
the derivation of a new exact solution describing plane waves. Additionally, a
scalar field potential, responsible for imparting mass to the bulk field, was
introduced in \cite{bb2}. Specifically, for an exponential scalar field
potential, new solutions were determined, including separable solutions and a
solution in which the scalar field is proportional to the logarithm of the
scale factor. Moreover, in \cite{bb3}, brane-world scalar field scenarios with
a four-dimensional de Sitter embedded space were investigated.

In this study, our focus lies on the symmetry analysis of the cosmological
field equations within brane-world gravity, specifically in an extended
spatially flat Friedmann--Lema\^{\i}tre--Robertson--Walker (FLRW) geometry
incorporating a bulk scalar field. The field equations in brane gravity
constitute a system of nonlinear partial differential equations, with an
unknown function representing the scalar field potential.

Symmetry analysis serves as a systematic approach to determine transformations
that preserve the invariance of the given dynamical system \cite{Bluman,
Stephani, olver}. When a symmetry vector exists, we can define a set of new
variables to reduce the given dynamical system through similarity
transformations. In the case of partial differential equations, applying
similarity transformations reduces the number of independent variables, while
in the case of ordinary differential equations, it leads to a dynamical system
of lower order. Additionally, symmetries can be employed to construct
conservation laws for the dynamical system \cite{ibra}.

The application of Lie symmetries in gravitational physics has found numerous
uses. In the context of the vacuum Einstein field equations within a
four-dimensional manifold, Lie symmetries were explored in \cite{ei0}, while a
similar analysis for a higher-dimensional manifold was presented in
\cite{ei1}. The inclusion of a perfect fluid was also considered in \cite{ei2}.

In cosmology, symmetry analysis has been widely employed to constrain the
unknown functions of the theory and construct conservation laws using
Noether's second theorem \cite{Bluman}. The derivation of conservation laws
has led to the discovery of new analytic and exact solutions for various
cosmological models, including those in scalar field theory \cite{ns1, ns2},
scalar-tensor theory \cite{ns3, ns4, ns5}, and modified theories of gravity
\cite{ns6, ns7}, as discussed in the review article \cite{ns0}. With this in
mind, the structure of the paper is organized accordingly.

n Section \ref{sec2}, we introduce the gravitational model and present the
field equations pertinent to our study, focusing on brane-world cosmology
within a five-dimensional manifold coupled with a bulk scalar field. Moving in
Section \ref{sec3} we solve the Lie symmetry classification problem by
determining all the functional forms for the scalar field potential where the
field equations admit Lie symmetries. We demonstrate the application of the
Lie symmetries and we construct similarity solutions for the field equations.
Finally, in Section \ref{sec4}, we summarize our findings and draw conclusions
from our analysis.

\section{Brane cosmology with bulk scalar field}

\label{sec2}

We consider a five-dimensional manifold with line element \cite{bb1}
\begin{equation}
ds^{2}=e^{2B\left(  t,y\right)  }\left(  -dt^{2}+dy^{2}\right)  +e^{2A\left(
t,y\right)  }\left(  dx_{1}^{2}+dx_{2}^{2}+dx_{3}^{2}\right)  , \label{le.01}%
\end{equation}
in which $A\left(  t,y\right)  $ and $B\left(  t,y\right)  $ are the two
scalar factors of the brane. In the line $y=const$, that is, $y=y_{0}$, the
line-element (\ref{le.01}) reduces to that of the FLRW, where $A\left(
t,y_{0}\right)  $ is the scale factor and $B\left(  t,y_{0}\right)  $ is the
lapse function, and $H\left(  t\right)  =e^{-B\left(  t,y_{0}\right)  }%
A_{,t}\left(  t,y_{0}\right)  $ is the Hubble function. The line element
(\ref{le.01}) admits a six-dimensional Killing algebra consisting of the
Killing symmetries of the three-dimensional Euclidean space.

Let $\phi$ be a bulk scalar field, we consider the gravitational Action
Integral to be \cite{bb2}
\begin{equation}
S=\int dx^{5}\sqrt{g}\left(  \frac{^{\left(  5\right)  }R}{2}-\frac{1}%
{2}\nabla_{\kappa}\phi\nabla^{\kappa}\phi-V\left(  \phi\right)  \right)  ,~
\end{equation}
where $\kappa=1,2,3,4,5$ and $^{\left(  5\right)  }R\,\,\ $is the Ricciscalar
for the five-dimensional metric (\ref{le.01})%
\begin{equation}
^{\left(  5\right)  }R=e^{-2B}\left(  B_{,tt}-B_{,yy}+3\left(  A_{,tt}%
+2A_{,t}^{2}-A_{,yy}-2A_{,y}^{2}\right)  \right)  .
\end{equation}

Thus, the gravitational field equations are%
\begin{equation}
^{\left(  5\right)  }G_{AB}=\left(  \nabla_{A}\phi\nabla^{A}\phi-g_{AB}\left(
\frac{1}{2}\nabla_{A}\phi\nabla^{A}\phi+V\left(  \phi\right)  \right)
\right)  , \label{le.02}%
\end{equation}
where $^{\left(  5\right)  }G_{AB}$ is the Einstein tensor for the line
element\ (\ref{le.01}). Furthermore, the equation of motion for the scalar
field, that is, the Klein-Gordon equation reads
\begin{equation}
\nabla^{2}\phi+V_{,\phi}=0. \label{le.03}%
\end{equation}
For the scalar field $\phi$, we assume that inherits the symmetries of the
background space, that is, $\phi=\phi\left(  t,x\right)  $.

We follow \cite{bb1,bb2}, thus we define the variables $\left(  u,v\right)  $
with transformation rule
\begin{equation}
u=t-y~,~v=t+y,
\end{equation}
that is%
\begin{equation}
t=\frac{u+v}{2}~,~y=\frac{v-y}{2}.
\end{equation}

Hence, in the new variables the line-element (\ref{le.01}) becomes%
\begin{equation}
ds^{2}=e^{2B\left(  u,v\right)  }\left(  -dudv\right)  +e^{2A\left(
u,v\right)  }\left(  dx_{1}^{2}+dx_{2}^{2}+dx_{3}^{2}\right)  , \label{le.001}%
\end{equation}
and the field equations (\ref{le.02}) are
\begin{align}
A_{,uv}+3A_{,u}A_{,v}  &  =\frac{1}{6}e^{2B}V\left(  \phi\right)
,\label{le.04}\\
-3A_{,uu}+6A_{,u}B_{,u}-3A_{,u}^{2}  &  =\phi_{,u}^{2},\label{le.05}\\
-3A_{,vv}+6A_{,v}B_{,v}-3A_{,v}^{2}  &  =\phi_{,v}^{2},\label{le.06}\\
B_{,uv}+2A_{,uv}+3A_{,u}A_{,v}  &  =-\frac{1}{2}\left(  \phi_{,u}\phi
_{,v}-\frac{1}{2}e^{2B}V\left(  \phi\right)  \right)  , \label{le.07}%
\end{align}
and the Klein-Gordon equation (\ref{le.03})
\begin{equation}
4\phi_{,uv}+6\left(  A_{,u}\phi_{,v}+A_{,v}\phi_{,u}\right)  +e^{2B}V_{,\phi
}=0. \label{le.08}%
\end{equation}

\section{Lie symmetries for the field equations}

\label{sec3}

In the augmented space $\left\{  u,v,A,B,\Phi\right\}  $ we consider the
infinitesimal transformation
\begin{align}
u^{\prime}  &  =u+\varepsilon\xi^{u}\left(  u,v,A,B,\phi\right)  ,\\
v^{\prime}  &  =v+\varepsilon\xi^{v}\left(  u,v,A,B,\phi\right)  ,\\
A^{\prime}  &  =A+\varepsilon\eta^{A}\left(  u,v,A,B,\phi\right)  ,\\
B^{\prime}  &  =B+\varepsilon\eta^{B}\left(  u,v,A,B,\phi\right)  ,\\
\phi^{\prime}  &  =\phi+\varepsilon\phi\left(  u,v,A,B,\phi\right)
\end{align}
with generator the vector field
\begin{equation}
\mathbf{X}=\xi^{u}\partial_{u}+\xi^{\nu}\partial_{v}+\eta^{A}\partial_{A}%
+\eta^{B}\partial_{B}+\eta^{\phi}\partial_{\phi}.
\end{equation}

Let $\mathbf{H}=0,~$describe the field equations (\ref{le.04})-(\ref{le.08}),
the $\mathbf{H}=0$ is invariant under the action of the latter infinitesimal
transformation if and only if there exist a function $\lambda$ such that the
following condition is satisfied%
\begin{equation}
\mathbf{X}^{\left[  2\right]  }\left(  \mathbf{H}\right)  =\lambda\mathbf{H},
\label{le.09}%
\end{equation}
or equivalently%
\begin{equation}
\mathbf{X}^{\left[  2\right]  }\left(  \mathbf{H}\right)  =0.
\end{equation}

$\mathbf{X}^{\left[  2\right]  }$ is the second extension/prolongation of $X$
in the jet space $\left\{  u,v,A,B,A_{,u},A_{,v},A_{uv},A_{uu},A_{,vv}%
,B_{,u},B_{,v},B_{uv},B_{uu},B_{,vv}\right\}  $ defined as
\begin{align}
\mathbf{X}^{\left[  2\right]  }  &  =\mathbf{X}+\eta_{\left[  u\right]  }%
^{A}\partial_{A_{,u}}+\eta_{\left[  v\right]  }^{A}\partial_{A_{,v}}%
+\eta_{\left[  u\right]  }^{B}\partial_{B_{,u}}\nonumber\\
&  +\eta_{\left[  v\right]  }^{B}\partial_{B_{,v}}+\eta_{\left[  u\right]
}^{\phi}\partial_{\phi_{,u}}+\eta_{\left[  v\right]  }^{\phi}\partial
_{\phi_{,v}}\\
&  \eta_{\left[  uu\right]  }^{A}\partial_{A_{,uu}}+\eta_{\left[  vv\right]
}^{A}\partial_{A_{,vv}}+\eta_{\left[  uv\right]  }^{A}\partial_{A_{,uv}%
}+\nonumber\\
&  \eta_{\left[  uu\right]  }^{B}\partial_{B_{,uu}}+\eta_{\left[  vv\right]
}^{B}\partial_{B_{,vv}}+\eta_{\left[  uv\right]  }^{B}\partial_{B_{,uv}}+\\
&  \eta_{\left[  uu\right]  }^{\phi}\partial_{\phi_{,uu}}+\eta_{\left[
vv\right]  }^{\phi}\partial_{\phi_{,vv}}+\eta_{\left[  uv\right]  }^{\phi
}\partial_{\phi_{,uv}},
\end{align}
with
\begin{equation}
\eta_{\left[  \alpha\right]  }^{A}=D_{\alpha}\eta^{A}-A_{,\beta}D_{\alpha}%
\xi^{\beta},
\end{equation}%
\begin{equation}
\eta_{\left[  \alpha\right]  }^{B}=D_{\alpha}\eta^{B}-B_{,\beta}D_{\alpha}%
\xi^{\beta},
\end{equation}%
\begin{equation}
\eta_{\left[  \alpha\right]  }^{\phi}=D_{\alpha}\eta^{\phi}-A_{,\beta
}D_{\alpha}\xi^{\beta},
\end{equation}%
\begin{equation}
\eta_{\left[  \alpha\beta\right]  }^{A}=D_{\alpha}\eta_{\left[  \beta\right]
}^{A}-A_{,\alpha\beta}D_{\alpha}\xi^{\beta},
\end{equation}%
\begin{equation}
\eta_{\left[  \alpha\beta\right]  }^{B}=D_{\alpha}\eta_{\left[  \beta\right]
}^{B}-B_{,\alpha\beta}D_{\alpha}\xi^{\beta},
\end{equation}%
\begin{equation}
\eta_{\left[  \alpha\beta\right]  }^{\phi}=D_{\alpha}\eta_{\left[
\beta\right]  }^{\phi}-\phi_{,\alpha\beta}D_{\alpha}\xi^{\beta},
\end{equation}
where $\alpha,\beta=u,v$ and we consider Einstein's summation convention.

The Lie symmetry condition (\ref{le.09}) applied to the field equations
(\ref{le.04})-(\ref{le.08}) leads to a system of partial differential
equations, from which the coefficients of the vector field $\mathbf{X}$ are
determined. The solution of these symmetry conditions is contingent upon the
functional form of the scalar field potential $V\left(  \phi\right)  $,
resulting in different allowed Lie symmetries for various potentials.

{From the Lie symmetry conditions it follows that the vector field }$X${
components }$\xi^{u},~\xi^{v},~\eta^{A},\eta^{B}${ and}$~\eta^{\phi}${ are }%
\[
\xi^{u}=\xi^{u}\left(  u\right)  ~,~\xi^{v}=\xi^{v}\left(  v\right)
~,~\eta^{A}=a_{0}%
\]%
\[
\eta^{B}=a_{1}\phi-\xi_{,u}^{u}-\xi_{,v}^{v}+a_{3}~,~\eta^{\phi}=3a_{1}%
\phi+\alpha_{2},
\]
{with constraint equations for the potential function }$V\left(  \phi\right)
${ and the free parameters }$a_{1}$\textbf{ and }$a_{2}$\textbf{,}%
\[
a_{1}AV_{,\phi}+a_{1}\phi V+a_{2}V_{,\phi}+2a_{3}V=0,
\]%
\[
7a_{2}V_{,\phi}+21a_{1}AV_{,\phi}+14a_{1}\phi-3a_{1}V_{,\phi}+14a_{3}V=0,
\]%
\[
3a_{1}AV_{,\phi}+a_{2}V_{,\phi\phi}+2a_{1}V+2a_{1}\phi V_{,\phi}%
+2a_{3}V_{,\phi}=0.
\]
\textbf{\qquad}

For arbitrary potential function the latter system admits a solution with
$a_{1}=0$, $a_{2}=0,~a_{3}=0$; that is, the allowed Lie symmetries are
\[
X_{1}=F\left(  u\right)  \partial_{u}-\frac{1}{2}F_{,u}\partial_{B}%
~,~X_{2}=G\left(  v\right)  \partial_{v}-\frac{1}{2}G_{,v}\partial_{B}%
~,~X_{3}=\partial_{A}.
\]
Vector fields $X_{1}$ and $X_{2}$ describe infinite number of Lie symmetry
vectors. The commutators of the allowed Lie symmetries are zero.

In the case where the potential function is the cosmological constant it
follows that $a_{1}=0$, $a_{3}=0$, i.e. $V\left(  \phi\right)  =\Lambda$, the
allowed Lie symmetries by the field equations (\ref{le.04})-(\ref{le.08}) are%
\[
X_{1}~,~X_{2}~,~X_{3}~,~X_{4}=\partial_{\phi}\text{,}%
\]
with zero commutators.

In the limit where $\Lambda=0$, that is, the scalar field is massless the
admitted symmetry vectors are
\[
X_{1}~,~X_{2}~,~X_{3}~,~X_{4}~,~X_{5}=\partial_{B}~,~X_{6}=\frac{1}{3}%
\phi\partial_{B}+A\partial_{\phi}\text{,}%
\]
with non-zero commutator $\left[  X_{3},X_{6}\right]  =X_{4}$.

Finally, for the exponential potential function $V\left(  \phi\right)
=V_{0}e^{\lambda\phi}$, we determine $a_{1}=0$,~$a_{3}=-\frac{\lambda}{2}%
a_{2}$; from where it follows that the admitted Lie symmetries for the
cosmological field equations (\ref{le.04})-(\ref{le.08}) are
\[
X_{1}~,~X_{2}~,~X_{3}\text{~and }X_{7}=-\frac{\lambda}{2}\partial_{B}%
+\partial_{\phi}\text{,}%
\]
where all the commutators are zero.

\subsection{Similarity solutions}

We proceed with the application of the Lie symmetries to reduce the field
equations and when it is feasible to determine new solutions.

Let $W=W\left(  u,v,A,B\right)  $ be a function and $X$ be a symmetry vector
for the function $W$, i.e. $X\left(  W\right)  =0$. Then, there we can always
define a new set of variables $\left\{  u,v,A,B\right\}  \rightarrow\left\{
\bar{u},\bar{v},\bar{A},\vec{B}\right\}  $, where the symmetry vector can be
written as $X=\partial_{\bar{u}}$, such that the symmetry condition becomes
$\frac{\partial W}{\partial\bar{u}}=0$, that is, $W=W\left(  \bar{v},\bar
{A},\bar{B}\right)  $. This transformation is called a similarity transformation.

In order to demonstrate the application of the Lie symmetries, in the
following lines we consider specific forms of the vector fields $X_{1}$ and
$X_{2}$ where closed-form similarity solutions are recovered.

\subsubsection{Cosmological solution}

Consider now the Lie symmetry vector $X_{1-2}^{0}=\partial_{u}-\partial_{v}$,
which is generated by the $X_{1},~X_{2}$ for $F\left(  u\right)  =1$ and
$G\left(  u\right)  =1$. The similarity transformation reads $\bar
{t}=u-v~,~\bar{y}=u+v$, $A=A\left(  \bar{t}\right)  $ and $B=B\left(  \bar
{t}\right)  $. In the new variables the line element (\ref{le.001}) becomes%
\begin{equation}
ds^{2}=e^{2B\left(  \bar{t}\right)  }\left(  -2d\bar{t}^{2}+2d\bar{y}%
^{2}\right)  +e^{2A\left(  \bar{t}\right)  }\left(  dx_{1}^{2}+dx_{2}%
^{2}+dx_{3}^{2}\right)  .
\end{equation}

The latter spacetime describes a five-dimensional homogeneous and anisotropic
geometry, the extension of the Bianchi I spacetime in the five dimensions with
two scale factors. The process of deriving the analytic solution for the field
equations, whether with or without the exponential potential, follows a
construction method similar to that presented in \cite{ns010} for the
four-dimensional Bianchi I spacetime.

Similarly, the vector field $X_{1-2}^{\prime0}=\partial_{u}+\partial_{v}$
induces a similarity transformation that characterizes a static and
anisotropic five-dimensional geometry with line element
\begin{equation}
ds^{2}=e^{2B\left(  \bar{y}\right)  }\left(  -2d\bar{t}^{2}+2d\bar{y}%
^{2}\right)  +e^{2A\left(  \bar{y}\right)  }\left(  dx_{1}^{2}+dx_{2}%
^{2}+dx_{3}^{2}\right)  .
\end{equation}

\subsubsection{Plane-wave solutions}

The application of Lie symmetries $X_{1}$ or $X_{2}$ yields plane-wave
solutions. However, it is worth noting that such solutions do not exist for a
nonzero potential \cite{bb2}. Conversely, vacuum plane-wave solutions have
been investigated in \cite{bb1}.

\subsubsection{Travel-wave solution}

Vector fields $X_{1}$ and $X_{2}$ correspond to an infinite number of
invariant transformations. Assume now that~$F\left(  u\right)  $ and $G\left(
v\right)  $ are quadratic functions, i.e. $F\left(  u\right)  =u^{2}$ and
$G\left(  v\right)  =v^{2}$, then $X_{1}^{2}=u^{2}\partial_{u}-u\partial_{B}$
and $X_{2}^{2}=v^{2}\partial_{v}-v\partial_{B}$. We consider the Lie symmetry
vector $X_{1-2}^{2}=X_{1}-X_{2}$ from where we derive the similarity
transformation%
\begin{equation}
A\left(  u,v\right)  =a\left(  w\right)  ~,~B=-\ln\left(  uv\right)  +b\left(
w\right)  ~,~\phi\left(  u,v\right)  =\psi\left(  w\right)  , \label{sd11}%
\end{equation}
where the new independent variable $w$ is defined as $w=\frac{u+u}{vu}$, or in
the $\left(  t,y\right)  $ coordinates, $w=\frac{2t}{t^{2}-y^{2}}$.

By replacing (\ref{sd11}) in the field equations (\ref{le.04})-(\ref{le.08})
we end with a system of nonlinear ordinary differential equations.

The reduced system is
\begin{align}
6\left(  a_{,ww}+3a_{,w}^{2}\right)  -V_{0}e^{2b+\lambda\psi}  &
=0,\label{dd.01}\\
3\left(  a_{,ww}+a_{,w}^{2}\right)  -6a_{,w}b_{,w}+\psi_{,w}^{2}  &  =0,\\
4\left(  \psi_{,ww}+3a_{,w}\psi_{,w}\right)  +V_{0}e^{2b+\lambda\psi}  &
=0,\\
4\left(  2a_{,ww}+3a_{,w}^{2}\right)  +4b_{,ww}+2\psi_{,w}^{2}+V_{0}%
e^{2b+\lambda\psi}  &  =0, \label{dd.04}%
\end{align}

For zero potential, $V_{0}=0$, the latter dynamical system admits the
closed-form solution%

\begin{equation}
A\left(  u,v\right)  =\ln\left(  f_{0}w^{p_{1}}\right)  ~,~B=-\ln\left(
uv\right)  +\ln\left(  b_{0}w^{p_{2}}\right)  ~,~\phi\left(  u,v\right)
=\frac{1}{\lambda}\ln\left(  \phi_{0}w^{p_{3}}\right)  ,
\end{equation}
with constraint equations
\[
p_{1}=\frac{1}{3}~,~p_{2}=\frac{3p_{3}^{2}-2\lambda^{2}}{6\lambda^{2}}\text{.}%
\]

For this exact solution the line element (\ref{le.001}) becomes%
\begin{equation}
ds^{2}=\left(  \frac{b_{0}w^{p_{2}}}{uv}\right)  ^{2}\left(  -dudv\right)
+\left(  f_{0}w^{\frac{1}{3}}\right)  ^{2}\left(  dx_{1}^{2}+dx_{2}^{2}%
+dx_{3}^{2}\right)  ,
\end{equation}
or%
\begin{equation}
ds^{2}=\left(  \frac{b_{0}}{\left(  t^{2}-y^{2}\right)  }\left(  \frac
{t^{2}-y^{2}}{2t}\right)  ^{p_{2}}\right)  ^{2}\left(  -dt^{2}+dy^{2}\right)
+\left(  f_{0}\left(  \frac{2t}{t^{2}-y^{2}}\right)  ^{\frac{1}{3}}\right)
^{2}\left(  dx_{1}^{2}+dx_{2}^{2}+dx_{3}^{2}\right)  .
\end{equation}

In the four-dimensional limit with $y=0$, the cosmological solution reads%
\begin{equation}
ds^{2}=\left(  \bar{b}_{0}\left(  t\right)  ^{p_{2}-1}\right)  ^{2}\left(
-dt^{2}\right)  +\left(  \bar{f}_{0}\left(  \frac{1}{t}\right)  ^{\frac{1}{3}%
}\right)  ^{2}\left(  dx_{1}^{2}+dx_{2}^{2}+dx_{3}^{2}\right)  .
\end{equation}

The scaling solution described represents a universe in which the cosmological
fluid behaves as an ideal gas with a constant equation of state parameter
$w_{\text{eff}}=2p_{2}-1$. For values of $p_{2}<\frac{1}{2}$, the universe
experiences acceleration.

However, for $V_{0}\neq0$, it is observed that the dynamical system does not
admit a real solution.

Similarly, we can utilize the Lie symmetry vectors $X_{1}$ and $X_{2}$ to
construct additional solutions in a similar manner.

\section{Conclusions}

\label{sec4}

In our study, we conducted an extensive classification of Lie symmetries for
the cosmological field equations in a five-dimensional spacetime with
brane-gravity featuring a bulk scalar field. By imposing the condition that
the field equations must possess non-trivial Lie symmetries, we derived the
scalar field potential. Through this approach, we determined the scalar field
potential in a manner that ensures the field equations admit Lie symmetries,
which form various Lie algebras.

For an arbitrary potential function, the field equations exhibit an infinite
number of Lie symmetries, forming the Lie algebra $G_{\infty}=2A_{\infty
}\oplus A_{1}$ \cite{pat1}. In the special case of a massless scalar field,
the admitted Lie symmetries give rise to the Lie algebra $G_{1}=2A_{\infty
}\oplus2A_{1}$. Furthermore, when the scalar field potential takes on a
constant value, representing the cosmological constant, the Lie symmetries
form the $G_{2}=2A_{\infty}\oplus A_{1}\oplus A_{2,1}$ Lie algebra.
Additionally, for the exponential potential, the Lie symmetries yield the
$G_{2}=2A_{\infty}\oplus2A_{1}$ Lie algebra.

n the last step of our research, we harnessed the Lie symmetries to establish
similarity transformations. These transformations played a crucial role in
reducing the field equations into ordinary differential equations and enabled
us to derive closed-form solutions. This application of Lie symmetries has
proven to be highly beneficial in the study of gravitational theories,
contributing valuable insights to the subject of symmetry analysis in this field.

Our findings underscore the usefulness and potency of the Lie analysis as a
method for exploring gravitational theories. Building on these results, we
intend to expand our research in the future by investigating the determination
of conservation laws for the gravitational field equations. This forthcoming
work will further enrich our understanding of the dynamics and conservation
principles within the realm of gravitational theories.

\textbf{Data Availability Statements:} Data sharing is not applicable to this
article as no datasets were generated or analyzed during the current study.

\begin{acknowledgments}
AP was partially financially supported by the National Research Foundation of
South Africa (Grant Numbers 131604). AP thanks the support of
Vicerrector\'{\i}a de Investigaci\'{o}n y Desarrollo Tecnol\'{o}gico (Vridt)
at Universidad Cat\'{o}lica del Norte through N\'{u}cleo de Investigaci\'{o}n
Geometr\'{\i}a Diferencial y Aplicaciones, Resoluci\'{o}n Vridt No - 098/2022.
\end{acknowledgments}


\begin{thebibliography}{99}                                                                                               %


\bibitem {ae1}A. Einstein, Grundgedanken der allgemeinen Relativit%
%TCIMACRO{\U{a8}}%
%BeginExpansion
\"{}%
%EndExpansion
atstheorie und Anwendung dieser Theorie in der Astronomie, Preussische
Akademie der Wissenschaften, Sitzungsberichte, 315, (1915)

\bibitem {grt1}B.P. Abbott et al., Observation of Gravitational Waves from a
Binary Black Hole Merger, Phys. Rev. Lett. 116, 061102 (2016)

\bibitem {grt2}M. Kramer et al., Strong-Field Gravity Tests with the Double
Pulsar, Phys.\ Rev. X 11, 041050 (2021)

\bibitem {grt3}C.M. Will, The Confrontation between General Relativity and
Experiment, Living Rev. Relativ. 17, 4 (2014)

\bibitem {rr1}A.G. Riess et al., Observational evidence from Supernovae for an
accelerating universe and cosmological constant, Astron. J. 116, 1009 (1998)

\bibitem {Teg}M. Tegmark et al., The Three-Dimensional Power Spectrum of
Galaxies from the Sloan Digital Sky Survey, Astrophys. J. 606, 702 (2004)

\bibitem {Kowal}M. Kowalski et al., Improved Cosmological Constraints from
New, Old and Combined Supernova Datasets, Astrophys. J. 686, 749 (2008)

\bibitem {conpr1}S. Weinberg, The cosmological constant problem, Rev. Mod.
Phys. 61, 1 (1989)

\bibitem {conpr2}T. Padmanabhan, Cosmological Constant - the Weight of the
Vacuum, Phys. Rept. 380, 235 (2003)

\bibitem {q4}P. Ratra and L. Peebles, Cosmological consequences of a rolling
homogeneous scalar field, Phys. Rev. D 37, 3406 (1988)

\bibitem {q11}T.P. Sotiriou, f(R) gravity and scalar--tensor theory, Class.
Quant. Grav. 23, 5117 (2006)

\bibitem {q1}J.C. Fabris, T.C. da C. Guio, M.H. Daouda and O.F. Piatella,
Scalar models for the generalized Chaplygin gas and the structure formation
constraints, Gravitation and Cosmology 17, 259 (2011)

\bibitem {br1}N. Arkani-Hamed, S. Dimopoulos and G. Dvali, The Hierarchy
Problem and New Dimensions at a Millimeter, Phys. Lett. B 429, 263 (1998)

\bibitem {br2}I. Antoniadis, N. Arkani-Hamed, S. Dimopoulos and G. Dvali, New
Dimensions at a Millimeter to a Fermi and Superstrings at a TeV, Phys. Lett. B
436, 257 (1998)

\bibitem {br3}L. Randall and R. Sundrum, Large Mass Hierarchy from a Small
Extra Dimension, Phys. Rev. Lett. 83, 3370 (1999)

\bibitem {br4}R.\ Maartens and K. Koyama, Brane-World Gravity, Living Rev.
Relat. 13, 5 (2010)

\bibitem {cc1}N. Arkani-Hamed, S Dimopoulos, N Kaloper and R. Sundrum, A Small
Cosmological Constant from a Large Extra Dimension, Phys. Lett. B 480, 193 (2000)

\bibitem {cc2}Z. Kakushadze, Bulk Supersymmetry and Brane Cosmological
Constant, Phys. Lett. B 489, 207 (2000)

\bibitem {cc3}S. Kachru, M. Schulz and E. Silverstein, Self-tuning flat domain
walls in 5d gravity and string theory, Phys. Rev. D 62, 045021 (2000)

\bibitem {cc4}S. Forste, Z. Lalak, S. Lavignac and H.P. Nilles, A Comment on
Self-Tuning and Vanishing Cosmological Constant in the Brane World, Phys.
Lett. B 481, 360 (2000)

\bibitem {bb0}O. DeWolfe, D.Z. Freedman, S.S. Gubser and A. Karch, Modeling
the fifth dimension with scalars and gravity, Phys.\ Rev. D 62, 046008 (2000)

\bibitem {bb01}O. Bertolami and C. Carvalho, Spontaneous symmetry breaking in
the bulk and the localization of fields on the brane, Phys. Rev. D 76, 104048 (2007)

\bibitem {bb1}G.\ T. Horowitz, I. Low and A. Zee, Self-tuning in an outgoing
brane wave model, Phys. Rev. D 62, 086005 (2000)

\bibitem {bb2}D. Langlois and M. Rodriguez-Martinez, Brane cosmology with a
bulk scalar field, Phys. Rev. D\ 64, 123507 (2001)

\bibitem {bb3}\'{E}.\'{E}. Flanagan, S.-H. Henry Tye and I. Wasserman, Brane
world models with bulk scalar fields, Phys. Lett. B 522, 155 (2001)

\bibitem {Bluman}G.W. Bluman and S. Kumei, Symmetries of Differential
Equations, Springer-Verlag, New York (1989)

\bibitem {Stephani}H. Stephani, Differential Equations: Their Solutions Using
Symmetry, Cambridge University Press, New York (1989)

\bibitem {olver}P.J. Olver, Applications of Lie Groups to Differential
Equations, Springer-Verlag, New York (1993)

\bibitem {ibra}N.H. Ibragimov, CRC Handbook of Lie Group Analysis of
Differential Equations, Volume I: Symmetries, Exact Solutions, and
Conservation Laws, CRS Press LLC, Florida (2000)

\bibitem {ei0}C.G. Torre and I.M. Anderson, Phys. Rev. Lett. 70, 23 (1993)

\bibitem {ei1}L. Marchildon, Lie Symmetries of Einstein's Vacuum Equations in
N Dimensions, J. Nonlin. Math. Phys. 5, 68 (1998)

\bibitem {ei2}H. Stephani, Symmetries of Einstein's field equations with a
perfect fluid source as examples of Lie--B\"{a}cklund symmetries, J. Math.
Phys. 29, 1650 (1988)

\bibitem {ns1}S. Basilakos, M. Tsamparlis and A. Paliathanasis, Using the
Noether symmetry approach to probe the nature of dark energy, Phys. Rev. D 83,
103512 (2011)

\bibitem {ns2}R. de Ritis, G. Marmo, G. Platania, C. Rubano and P. Scudellaro,
New approach to find exact solutions for cosmological models with a scalar
field, Phys. Rev. D 42, 1091 (1990)

\bibitem {ns3}A. Paliathanasis, M. Tsamparlis, G. Leon and A. Paliathanasis,
New Conservation Laws and Exact Cosmological Solutions in Brans--Dicke
Cosmology with an Extra Scalar Field, Symmetry 13, 1364 (2021)

\bibitem {ns4}P. A. Terzis, N. Dimakis and T. Christodoulakis, Noether
analysis of scalar-tensor cosmology, Phys. Rev. D 90, 123543 (2014)

\bibitem {ns5}A. Paliathanasis, M. Tsamparlis, S.\ Basilakos and S.
Capozziello, Scalar-Tensor Gravity Cosmology: Noether symmetries and
analytical solutions, Phys. Rev. D 89, 063532 (2014)

\bibitem {ns6}S. Capozziello, K. Dialektopoulos and S.V. Sushkov,
Classification of the Horndeski cosmologies via Noether symmetries, Eur. Phys.
J. C 78, 447 (2018)

\bibitem {ns7}N. Sk., Noether symmetry in f(T) teleparallel gravity, Phys.
Lett. B 775, 100 (2017)

\bibitem {ns0}M. Tsamparlis and A.\ Paliathanasis, Symmetries of Differential
Equations in Cosmology, Symmetry 10, 233 (2018)

\bibitem {ns010}M. Tsamparlis and A. Paliathanasis, The geometric nature of
Lie and Noether symmetries, Gen. Relat. Grav. 43, 1861 (2011)

\bibitem {pat1}J. Patera, R.T. Sharp, P. Wintenritz and H.\ Zassenhaus,
Invariants of real low dimension Lie algebras, J. Math. Phys. 17, 986 (1976)
\end{thebibliography}
\end{document}